\documentclass{du-journals}
\journal{Journal of Holography Applications in Physics}
\ArticleType{Regular article}
\Year{Spring 2025}
\Vol{5}
\No{2}
\Page{\thepage--\pageref{LastPage}}
\Received{January xx, 2025}
\Revised{January xx, 2025}
\Accepted{January xx, 2025}
\Doihead{10.22128/jhap.2021.452.xxxx}
\title{$\kappa$-Entanglement Entropy from Black Hole}
\subtitle{$\kappa$-Entanglement Entropy from Black Hole}
\author[1]{{Fabiano F. Santos}}
\address[1]{School of Physics, Damghan University, Damghan, 36716--41167, Iran. \\Departamento de Física, Universidade Federal do Maranhão, São Luís, 65080-805, Brazil.;\\ Corresponding Author E-mail: fabiano.ffs23@gmail.com}
\author[2]{{Ankit Anand}}
\address[2]{Department of Physics, Indian Institute of Technology, Kanpur 208016, India; \\ E-mail: anand@iitk.ac.in}
\begin{document}

\begin{abstract}
In this work, we explore the entanglement entropy equipped with the $\kappa$-algebra. This entanglement entropy is computed through the geometric setup as performed by Hartman-Maldacena, which, in their prescription, finds that the entropy grows linearly in time. In our case, we show that the $\kappa$-algebra embedding provides a richer scenario where the third-order corrections in time added from $\kappa$-algebra to entanglement entropy imply that the growth of quantum correlations between subsystems is more intricate than a simple linear increase into the dynamics of black hole thermalization and quantum information flow. In the context of holography, such corrections suggest that the thermalization process is not instantaneous but involves higher-order interactions between subsystems.
\end{abstract}


\newpage

\tableofcontents

\newpage

\section{Introduction}

The Kaniadakis algebra, or $\kappa$-algebra, is a generalization of classical statistical mechanics that introduces a deformation parameter $\kappa$, allowing for the description of systems with non-extensive or relativistic properties \cite{Kaniadakis:1997aq, Kaniadakis:2002ie, Kaniadakis:2004rh, Kaniadakis:2024oro}. This framework has been applied in various fields, including cosmology, black hole thermodynamics, and quantum mechanics \cite{Nojiri:2022aof,Nojiri:2022dkr,Nojiri:2023bom,Nojiri:2023ikl,Luciano:2024bco, Ambrosio:2024dtk, Yarahmadi:2024lzd, Lehmann:2024kce}. Entanglement entropy, on the other hand, quantifies the degree of quantum entanglement between subsystems and plays a crucial role in understanding quantum field theory, black hole physics, and holography \cite{Ambrosio:2024dtk}.

Recent advancements in theoretical physics indicate that the Kaniadakis entropy, rooted in the framework of $\kappa$-statistics, offers promising insights into the nature of entanglement entropy, particularly in systems exhibiting non-extensive or relativistic dynamics \cite{Ambrosio:2024dtk, Santos:2022fbq, Santos:2020txg}. This perspective aligns with the computation of entanglement entropy through the AdS/CFT correspondence \cite{Ryu:2006bv,Santos:2024cvx, DosSantos:2022exb,Caceres:2017lbr}, a powerful connection between gravitational theories in anti-de Sitter (AdS) spaces with conformal field theories (CFTs) on their boundaries. A notable extension of this framework is the AdS/BCFT correspondence, where the boundary conformal field theory (BCFT) is defined on a manifold with a physical boundary \cite{Santos:2024cvx,Santos:2021orr,Santos:2023mee, Santos:2024cwf,Takayanagi:2011zk, Fujita:2011fp,Kanda:2023zse}. In this scenario, the bulk spacetime is characterized by an asymptotically AdS geometry that terminates on an end-of-the-world (EoW) brane \cite{Maldacena:1997re, Witten:1998qj, 2840490}. The EoW brane plays a crucial role in encoding the boundary conditions of the BCFT, thereby enriching our understanding of the interplay between bulk and boundary physics.

In the AdS/CFT framework, the computation of entanglement entropy is elegantly achieved through the Ryu-Takayanagi (RT) prescription \cite{Ryu:2006bv} or its dynamic extension, the Hubeny-Rangamani-Takayanagi (HRT) formula \cite{Santos:2024cvx, DosSantos:2022exb, Caceres:2017lbr}. To illustrate the RT prescription, consider a static bulk spacetime $\mathcal{M}$ foliated as $\mathcal{M}=\Pi_{t}\mathcal{N}_{t}\times\mathcal{R}_{t}$, with its boundary defined as $\partial\mathcal{M}=\Pi_{t}\partial\mathcal{N}_{t}\times\mathcal{R}_{t}$. By partitioning the boundary $\partial\mathcal{N}_{t}$ into two complementary regions, 
$\mathcal{A}$ and $\mathcal{B}$, we can define an entangling surface $\partial\mathcal{A}$ that separates these regions. The Holographic Entanglement Entropy (HEE) associated with this surface is then determined by the minimal area of a bulk surface anchored to $\partial\mathcal{A}$, as prescribed by the RT formula. This approach provides a profound connection between quantum entanglement in the boundary theory and the geometry of the bulk spacetime. The Holographic Entanglement Entropy (HEE) across this entangling surface $\partial\mathcal{A}$ is given by
\begin{eqnarray}
S(\mathcal{A})=\frac{A(\gamma_{\mathcal{A}})}{4G}.\label{EET}  
\end{eqnarray}
Here $\gamma_{\mathcal{A}}$ is the codimensional-2 minimal surface in the bulk; It ends on $\mathcal{A}$: $\partial\gamma_{\mathcal{A}}\gamma_{\mathcal{A}}|_{\partial\mathcal{M}}=\partial\mathcal{A}$. It is homologous to $\mathcal{A}$: $\exists\,\mathcal{R}\subset\mathcal{M}\,:\,\partial\mathcal{R}=\gamma_{\mathcal{A}}\cup\mathcal{A}$ with $\mathcal{R}$ smooth, interpolating codimension-1 surface (homology constraint) \cite{Ryu:2006bv}.

\begin{figure}[!ht]
\begin{center}
\includegraphics[width=\textwidth]{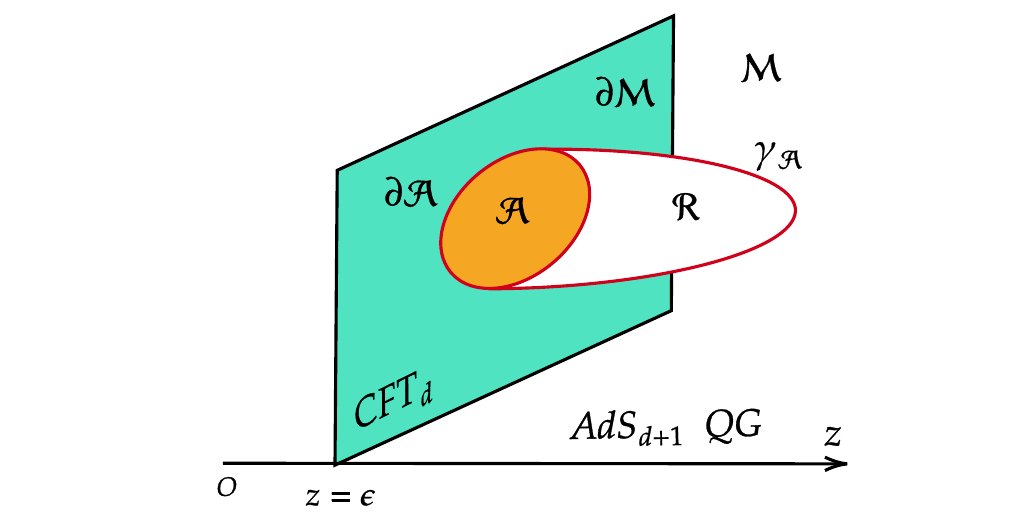}
\end{center}
\caption{Shcematic Ryu-Takayanagi formula. The Holographic Entanglement Entropy (HEE) of the subregion $\mathcal{A}$ on the boundary $CFT$ can be computed through the area of the minimal surface $\gamma_{\mathcal{A}}$ extending into the bulk.}\label{HH}
\label{planohwkhz}
\end{figure}

The conjecture presented in Eq. (\ref{EET})-\cite{Ryu:2006bv} establishes a profound connection between the entanglement entropy of a region 
$\mathcal{A}$ in a conformal field theory (CFT) with a holographic dual and the geometry of a minimal surface in the bulk spacetime. Specifically, the entanglement entropy ($S$) is determined by the area of a minimal surface ($\mathcal{A}$) that extends from the conformal boundary \cite{Santos:2024cvx, Geng:2021iyq, Geng:2024xpj,Geng:2022dua, Geng:2020fxl}, where $\mathcal{A}$ resides, into the bulk volume, as illustrated in Fig. \ref{HH}. This reduces the problem of minimizing a classical area functional, making it a purely geometric computation. In the context of boundary conformal field theory (BCFT), the minimal surface is further constrained to intersect the end-of-the-world (EoW) brane, with the entanglement entropy of $\mathcal{A}$ being proportional to the area of this surface. This framework highlights the deep interplay between quantum entanglement in the boundary theory and the geometry of the bulk spacetime.

Recently, new ways to investigate the black hole information have emerged, for example, entanglement islands and the Page curve  \cite{Geng:2021iyq, Geng:2024xpj,Geng:2022dua,Geng:2020fxl,Geng:2024xpj} within the framework on a Karch-Randall braneworld background \cite{Karch:2000ct, DeWolfe:2001pq, Santos:2023eqp, Brito:2018pwe}. By analytically examining the holographic boundary conformal field theory, the Horndeski parameters introduce significant deviations in the Page curve compared to predictions from standard general relativity \cite{Santos:2024cvx}. These deviations arise from the nontrivial geometric effects induced by the Horndeski scalar field, which fundamentally alter the bulk spacetime structure. Notably, these findings reveal that the geometry far from the AdS limit plays a more prominent role than previously recognized, underscoring the impact of Horndeski gravity on the distribution of quantum information in holographic models \cite{Santos:2024cvx}. 

In this work, we propose an investigation using the $\kappa$-algebra to probe the effects arising from the geometry modified by this algebra on the Hartman-Maldacena entanglement entropy \cite{Hartman:2013qma}. In their study, Hartman and Maldacena delve into the intricate and complex nature of black holes, focusing on the significant challenges involved in establishing a theoretical connection between the black hole's interior and its exterior. The interior of a black hole, often referred to as the "future region," is a domain where the laws of physics begin to break down as we understand them due to the extreme gravitational forces and spacetime curvature. This region is hidden behind the event horizon, a boundary beyond which no information or matter can escape to the outside universe. On the other hand, the exterior of the black hole represents the observable universe, where physical processes can be studied and measured. The difficulty in connecting these two regions arises from the fundamental limitations imposed by general relativity and quantum mechanics, as well as the information paradox, which questions how information about matter falling into a black hole can be preserved or retrieved. Hartman and Maldacena's work sheds light on these challenges, offering insights into the interplay between gravity, quantum theory, and the nature of spacetime itself. Furthermore, extremal surfaces do not penetrate the event horizon in the static case. However, for a time-dependent system, we can probe its interior. Using black brane geometry, we will probe the impact of $\kappa$-algebra on the evolution of entanglement entropy. For this, we will consider that the functions for the geometry embedded as a codimension sub-manifold of a Minkowski space are generalized $\kappa$-functions. Thus, in the framework of Kaniadakis \cite{Kaniadakis:1997aq, Kaniadakis:2002ie, Kaniadakis:2004rh, Kaniadakis:2024oro}, the mean idea is the analysis of how $\kappa$-deformed statistics modifies the traditional understanding of entropy. With this, we open new avenues with potential implications for entanglement entropy in relativistic and non-extensive systems \cite{deSousa:2025gex}.

\section{Review of $\kappa-$ deformed Theory}\label{Sec:Review of kappa deformed Theory}

The $\kappa$-deformed formalism is based on generalizing the standard exponential and logarithm functions. The $\kappa$-exponential is defined by
\begin{eqnarray}\label{Kappa deformed Exp}
\exp_\kappa{u} &\equiv& \left( \kappa u + \sqrt{1+\kappa^2 u^2} \right)^{1/\kappa} = \exp \left( \frac{1}{\kappa} \textrm{Arcsinh} (\kappa u) \right) \ .
\end{eqnarray}

\quad The $\kappa$-logarithm is
\begin{eqnarray}\label{Kappa deformed Log}
 \displaystyle \log_\kappa {u} &\equiv& \frac{u^\kappa - u^{-\kappa}}{2\kappa} = \frac{1}{\kappa} \sinh (\kappa \log u) \ .
\end{eqnarray}
The \(\kappa\)-exponential and \(\kappa\)-logarithm functions satisfy the following fundamental algebraic properties
\begin{eqnarray}
\exp_\kappa (a) \cdot \exp_\kappa (b) = \exp_\kappa (a \overset{\kappa}{\oplus} b) \;\;\;\;\;&;&\;\;\;\;\; \frac{\exp_\kappa (a)}{\exp_\kappa (b)} = \exp_\kappa (a \overset{\kappa}{\ominus} b) \nonumber \\
\log_\kappa (a b) =  \log_\kappa (a) \overset{\kappa}{\oplus} \log_\kappa (b)  \;\;\;\;\;&;&\;\;\;\;\;  \log_\kappa \left(\frac{a}{b}\right) =  \log_\kappa (a) \overset{\kappa}{\ominus} \log_\kappa (b) \ . \nonumber 
\end{eqnarray}
Here, the \(\kappa\)-deformed addition operator, denoted by \( \overset{\kappa}{\oplus} \) and subtraction operation, \( \overset{\kappa}{\ominus} \) respectively, are
\begin{equation}
a \overset{\kappa}{\oplus} b = a \sqrt{1+\kappa^2 b^2} + b \sqrt{1+\kappa^2 a^2}  \;\;\;\;\;;\;\;\;\;\; a \overset{\kappa}{\ominus} b = a \sqrt{1+\kappa^2 b^2} - b \sqrt{1+\kappa^2 a^2} \ .
\end{equation}  
These identities establish the fundamental structural relationships governing the \(\kappa\)-exponential and \(\kappa\)-logarithm functions within the framework of \(\kappa\)-deformed algebra. Such modifications to standard mathematical operations arise naturally in generalized statistical and thermodynamical formalisms \cite{KANIADAKIS2001405, Scarfone-2015}. Using the definitions of the $\kappa$-deformed exponential function as in Eq.~\eqref{Kappa deformed Exp} and the $\kappa$-deformed logarithm in Eq.~\eqref{Kappa deformed Log}, by expanding, one can observe in the limit $\kappa \rightarrow 0$ results in standard exponential function as 
\begin{eqnarray}
\exp_\kappa{u} &=& e^u-\frac{1}{6} \kappa ^2 \left(e^u u^3\right)+\frac{1}{360} \kappa ^4 e^u u^5 (5 u+27)+ \mathcal{O}\left(\kappa ^5\right) \ , \nonumber \\
 \displaystyle \log_\kappa {u} &=& \log (u)+\frac{1}{6} \kappa ^2 \log ^3(u)+\frac{1}{120} \kappa ^4 \log ^5(u)+O\left(\kappa ^5\right) \ . \nonumber
\end{eqnarray}
Now, using this, we can easily compute the trigonometric and hyperbolic functions as 
\begin{eqnarray}
    \sin_\kappa{x} &=& \frac{\left(\exp_\kappa{(i\,x)}\right)-\left(\exp_\kappa{(-i\,x)}\right)}{2 i} =\sin (x )+\frac{1}{6} \kappa ^2 x ^3 \cos (x )+\mathcal{O}\left(\kappa ^3\right)\nonumber \\
     \cos_\kappa{x} &=& \frac{\left(\exp_\kappa{(i\,x)}\right)+\left(\exp_\kappa{(-i\,x)}\right)}{2} = \cos (x )-\frac{1}{6} \kappa ^2\,x ^3 \sin (x )+\mathcal{O}\left(\kappa ^3\right)\nonumber \\
      \sinh_\kappa{x} &=& \frac{\left(\exp_\kappa{(x)}\right)-\left(\exp_\kappa{(-x)}\right)}{2} = \sinh (x )-\frac{1}{6} \kappa ^2 \,x ^3 \cosh (x )+\mathcal{O}\left(\kappa ^3\right) \nonumber \\
       \cosh_\kappa{x} &=& \frac{\left(\exp_\kappa{(x)}\right)+\left(\exp_\kappa{(-x)}\right)}{2} = \cosh (x )-\frac{1}{6} \kappa ^2 \, x ^3 \sinh (x ) + \mathcal{O}\left(\kappa ^3\right) \ . \nonumber 
\end{eqnarray}
Again, one returns to undeformed theory in the limit $\kappa \rightarrow 0$.
\subsubsection*{$\kappa$-deformed Calculus}

A $\kappa$-deformed calculus can be naturally introduced such that it reduces to conventional calculus in the limit \( \kappa \to 0 \). The $\kappa$-derivative is defined as  
\begin{equation}
D_{\kappa} f(x) = \frac{d f(x)}{d x_{\kappa}} = \lim_{y \to x} \frac{f(x) - f(y)}{x \overset{\kappa}{\ominus} y} \ ,
\end{equation}
where the $\kappa$-differential \( dx_{\kappa} \) is given by  
\begin{equation}
dx_{\kappa} = \lim_{y \to x} (x \overset{\kappa}{\ominus} y) = \lim_{y \to x} \frac{x^2 - y^2}{x \overset{\kappa}{\oplus} y} = \frac{dx}{\sqrt{1 + \kappa^2 x^2}} \ . 
\end{equation}
Integrating this expression yields the function  
\begin{equation}
x_{\kappa} = \frac{1}{\kappa} \ln \left( \sqrt{1 + \kappa^2 x^2} + \kappa x \right) = \frac{1}{\kappa} \operatorname{arcsinh}(\kappa x) \ ,
\end{equation}
and easy to see it satisfies the relation  
\begin{equation}
x_{\kappa} + y_{\kappa} = (x \overset{\kappa}{\oplus} y)_{\kappa} \ .
\end{equation}
Furthermore, the $\kappa$-derivative can be rewritten in the form  
\begin{equation}
\frac{d f(x)}{d x_{\kappa}} = \sqrt{1 + \kappa^2 x^2} \frac{d f(x)}{dx} \ .
\end{equation}
The $\kappa$-deformed exponential and hyperbolic functions satisfy the differential relations  
\begin{eqnarray}
\frac{d}{d x_{\kappa}} \exp_{\kappa}(x) = \exp_{\kappa}(x) \;\;\;\;\;\ ,\;\;\;\;\; \frac{d}{d x_{\kappa}} \sinh_{\kappa}(x) = \cosh_{\kappa}(x).
\end{eqnarray}

Finally, using the definition of the $\kappa$-derivative, the corresponding $\kappa$-deformed integral is defined as 
\begin{equation}
\int f(x) \, d x_{\kappa} = \int \frac{f(x)}{\sqrt{1 + \kappa^2 x^2}} \, dx.
\end{equation}
This formulation ensures that the fundamental principles of integral calculus remain valid in the $\kappa$-deformed framework. Specifically, the linearity, additivity, and fundamental theorem of calculus continue to hold, with the $\kappa$-integral acting as the inverse operation of the $\kappa$-derivative. The deformation parameter \( \kappa \) introduces a non-trivial weighting factor \( \frac{1}{\sqrt{1 + \kappa^2 x^2}} \), which modifies the measure of integration while preserving the core structure of conventional integral calculus.



\section{Hartman-Maldacena entanglement entropy}\label{Sec:Hartman-Maldacena entanglement entropy}

\begin{figure}[!ht]
\begin{center}
\includegraphics[width=\textwidth]{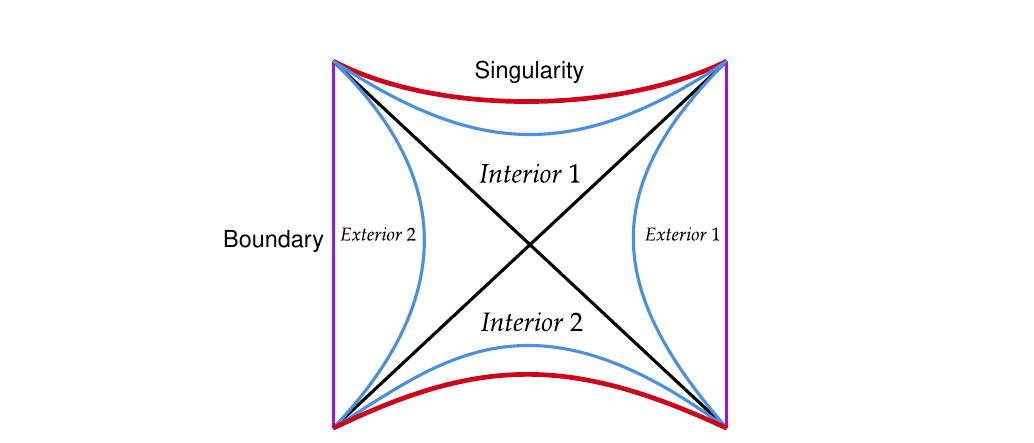}
\end{center}
\caption{The Penrose diagram for the maximally extended black branes under consideration is presented, with the spatial coordinates along the brane suppressed for simplicity. The diagram features two distinct exterior regions, denoted as $Exterior\,1$ (right side) and $Exterior\,2$ (Left side), each associated with a boundary where the corresponding dual field theories reside. Additionally, the diagram includes two interior regions: one in the future, labeled $Interior\,1$, and another in the past, labeled 
$Interior\,2$. This configuration can be constructed from an Euclidean solution by employing a method of analytic continuation, specifically by joining the solution across a moment of time-reflection symmetry.}\label{HW}
\label{planohwkhz}
\end{figure}

In this section, we present the general spatial embedding formalism to compute the Hartman-Maldacena (HM)
entanglement entropy \cite{Hartman:2013qma}. Considering the black-string insight  
\begin{eqnarray}
ds^{2}_{AdS_4}=\frac{1}{r^{2}\sin^{2}_{\kappa}(u)}\left(-f(r)dt^{2}+dy^2+r^{2}du^{2}
+\frac{dr^{2}}{f(r)}\right) \label{me}  
\end{eqnarray}
The interior region corresponds to $t\to\,t_{int.}-\frac{i\beta}{2}$ Fig \ref{HW} and $r\,e^{t}$
is finite as we cross the horizon, and $t_{int}$ is real. The coordinate $t_{int.}$ is spacelike in the interior. 
For the metric (\ref{me}), because a dual field theory is a CFT$_3$ on an AdS$_3$ black 
hole background with conformal boundary conditions at $r=0$, we obtain
\begin{eqnarray}
ds^{2}_{AdS_3}=\frac{1}{r^{2}}\left(-f(r)dt^{2}+dy^{2}+\frac{dr^{2}}{f(r)}
\right)\,.
\end{eqnarray}
To simplify our results, we rewrite $f(r)$ as 
\begin{equation}
f(r)=1-\frac{r^2}{r^{2}_{h}},\label{med2}
\end{equation}
Entanglement entropy serves as a fundamental tool in the study of holography, where it is computed using the area of an extremal surface in the AdS spacetime that terminates on the boundary \cite{Ryu:2006bv, Hubeny:2007xt, Nishioka:2009un}. In static configurations, these extremal surfaces remain confined outside the event horizon \cite{Hubeny:2012ry}. However, in dynamical, time-dependent scenarios, they can extend beyond the horizon, providing a unique means to probe the interior structure of the spacetime. This distinction underscores the critical role of time dependence in accessing deeper insights into the geometry and physics of black holes. For this, we consider the following geometry embedded as a codimension-one sub-manifold of a four-dimensional Minkowski space:
\begin{eqnarray}
ds^2=\eta_{ab}dX^{a}dX^{b};\,\eta_{ab}=diag(1,1,-1,-1),
\end{eqnarray}
where the embedding equation is $X_{a}X^{a}=1$. with a re-parameterization like $sin^{-1}(u)=\cosh(\rho)$

\begin{equation}
ds^{2}_{AdS_3}=\frac{\cosh_{\kappa}(\rho)}{r^{2}}\left(-f(r)dt^{2}+\frac{dr^{2}}{f(r)}
\right)+d\rho^2.\label{med1}
\end{equation}

By adopting the following parameterization in the $kappa-$deformation for the embedding equation 
\begin{eqnarray}\label{Embeddings}
&&X_{0}=\frac{2r_h-r}{r}\cosh_\kappa(\rho), \nonumber \\  
&&X_{1}=\frac{2r_h}{r}\sqrt{\left(1-\frac{r^2}{r^{2}_{h}}\right)}  
\sin_\kappa\left(\frac{2\pi\,t}{\beta}\right)\cosh_\kappa(\rho),\nonumber \\  
&&X_{2}=\frac{2r_h}{r}\sqrt{\left(1-\frac{r^2}{r^{2}_{h}}\right)}  
\cos_\kappa\left(\frac{2\pi\,t}{\beta}\right)\cosh_\kappa(\rho), \nonumber \\  
&&X_{3}=\sinh_\kappa(\rho) \ ,  
\end{eqnarray}  
where the inverse Hawking temperature is given by \( \beta=4\pi\,r_h=1/T \), we can efficiently compute the Hartman-Maldacena surface area without explicitly solving the geodesic differential equation associated with the minimal surface. Within this embedding framework, the geodesic length \( l \) can be determined using the coordinates \((X_0, X_1, X_2, X_3)\) and their corresponding primed counterparts \((X'_0, X'_1, X'_2, X'_3)\) via the relation
\begin{eqnarray}  
l=\cosh^{-1}(X_0X'_0 + X_1X'_1 - X_2X'_2 - X_3X'_3) \ .  
\end{eqnarray} 
Unlike the island surface, which extends from the bipartition to the KR brane, the Hartman-Maldacena (HM) surface passes over an Einstein-Rosen bridge before terminating at the right bipartition within the right-side thermofield double \cite{Hartman:2013qma}.  The left and right bipartitions of the \((u, \rho)\) coordinate system are placed at \((u, \rho) = (u_L, 1)\) and \((u, \rho) = (u_R, 1)\), respectively \cite{Santos:2024cvx}.  To use the embedding given in equation \eqref{Embeddings}, we add a regularization parameter \(\rho_{\epsilon} \), which we finally take to infinity to ensure that the bipartitions dwell on the asymptotic boundary. The time coordinate transformation for the right bipartition is as follows: $t \rightarrow -t+i\frac{\beta}{2}$. This corresponds to reversing the time-like Killing vector field on the opposite side of the black hole.
The area of the Hartman-Maldacena surface may now be easily computed using equation \eqref{Embeddings} as 
\begin{eqnarray}\label{Def AHM}
&&\!\!\!\!\!\!\!\!\!\!\!\!\!\!\! \!
\mathcal{A}_{HM} = \cosh^{-1}(X_0^LX^R_0 + X^L_1X^R_1 - X^L_2X^R_2 - X^L_3X^R_3) = \cosh^{-1} (Z_0+\kappa^2 Z_1) \nonumber \\
&& = \cosh^{-1}\left[Z_0\right] + \kappa^2 \frac{Z_1}{\sqrt{Z_0^2-1}}  \ , 
\end{eqnarray}
where in the terms of $\Delta_L=r_h-r_L,\,\Delta_R=r_h-r_R$ 
\begin{eqnarray}
    Z_0 &=& \frac{(\Delta_L+r_h) (\Delta_R+r_h)+4 r_h \omega  \sqrt{\Delta_L \Delta_R} \cosh \left(\frac{2 \pi  t}{\beta }\right)}{(r_h-\Delta_L) (r_h-\Delta_R)} \cosh{\rho} -\sinh^2{\rho} \nonumber \\
    Z_1 &=& \frac{r_h \left(\beta ^3 \rho ^3 (\Delta_L+\Delta_R) \sinh (2 \rho )+2\omega \sqrt{\Delta_L \Delta_R } \right)}{3 \beta ^3 (r_h-\Delta_L) (r_h-\Delta_R)} \Bigg\{-\beta ^3 \rho ^3 \sinh (2 \rho ) \cosh \left(\frac{2 \pi  t}{\beta }\right) \nonumber \\ 
    && \;\;\;\;\;\;\;\;\;\;\;\;\;\;\;\;\;\;\;\;\;\;\;\;\;\;\;\;\;\;\;\;\;\;\;\;\;\;\;\;\;\;\;\;\;\;\;\;\;\;\;\;\;\;\;\;\; +(16-24 i) \pi ^3 t^3 \cosh ^2(\rho ) \sinh \left(\frac{2 \pi  t}{\beta }\right)\Bigg\} \ .\nonumber 
\end{eqnarray}
By employing hyperbolic trigonometric identities and using Eq.\eqref{Def AHM}, we derive the following expression
\begin{eqnarray}
S_{HM} &=& \frac{\mathcal{A}_{HM}}{4G}= \frac{c}{6} \log \left[\frac{r_h}{r_L r_R} \left(\Delta_L + \Delta_R + 2\omega(\rho_{\epsilon})\sqrt{\Delta_L \Delta_R} \cosh\left(\frac{4\pi t}{\beta}\right)\right) \right] + \frac{c}{3} \rho_{\epsilon} + S^{\kappa}_{HM} \ , \nonumber  \label{SHM}
\end{eqnarray}
with
\begin{eqnarray}
S^{\kappa}_{HM}=\frac{c}{6}\kappa^2 \frac{Z_1}{\sqrt{Z_0^2-1}}\label{SHMK}
\end{eqnarray}
where \( c = \frac{3}{2G} \) represents the central charge-like quantity \cite{Santos:2024cvx, Geng:2022dua,DosSantos:2022exb}, and it is easy to verify 
\begin{eqnarray}
    \frac{Z_1}{Z_0} = \frac{e^{2 \rho } r_h \left(-\beta ^3 \rho ^3 (\Delta_L+\Delta_R)+2r_h \omega \sqrt{\Delta_L \Delta_R}  \left(\beta ^3 \rho ^3 \cosh \left(\frac{2 \pi  t}{\beta }\right)-(8-12 i) \pi ^3 t^3 \sinh \left(\frac{2 \pi  t}{\beta }\right)\right)\right)}{3 \beta ^3 \left(2 (r_h-\Delta_L) (r_h-\Delta_R)+e^{2 \rho } r_h \left(\Delta_L+\Delta_R-2\omega \sqrt{\Delta_L \Delta_R}   \cosh \left(\frac{2 \pi  t}{\beta }\right)\right)\right)} \ . \nonumber
\end{eqnarray}  
It is evident that in the limit \( \kappa \to 0 \), the result seamlessly recovers the corresponding expression from \cite{Santos:2024cvx, Geng:2022dua}. In the next section, we study the evolution of the entanglement entropy. We can see that the corrections in $\kappa^2$ for equation (\ref{SHM}) provide thermalization process is non-instantaneous, which involves higher-order interactions between subsystems.
\section{The growth of the entanglement entropy}
In the prescription of \cite{Santos:2024cvx, Geng:2021iyq, Geng:2024xpj, Geng:2020fxl, Hartman:2013qma}, the black hole corresponds to the holographic duals of thermal field theories. In our work, we set the temperature to $\beta=1/T$.

The extended Penrose diagram includes a second exterior region Fig \ref{HW}, mirroring the first, which plays a crucial role in the dual description of the full spacetime. This spacetime is holographically dual to the thermofield double state, as described in prior work \cite{Hartman:2013qma}. The duality involves two distinct boundaries, each corresponding to a separate copy of the field theory. These two field theory copies are entangled in a thermofield double state, which can be expressed as a superposition over energy eigenstates. Notably, when the field theories are defined on a non-compact spatial manifold, the index $n$ in the entangled state becomes continuous, reflecting the unbounded nature of the spectrum.
\begin{eqnarray}
|\Psi\rangle = \sum_{n} |E_n \rangle_1 |E_n\rangle_2 e^{ - { \beta \over 2 } E_n } \ .   
\end{eqnarray}

We now turn our attention to the entanglement entropy of a specific region, denoted as "$\mathcal{A}$". This region is defined as comprising one-half of the spatial domain in each of the two copies of the thermofield double state \cite{Santos:2024cvx,Geng:2021iyq,Geng:2024xpj,Geng:2020fxl}. To analyze this, we partition each copy into two equal halves at a fixed time $t_b$, which is synchronized across both boundaries. Importantly, we consider the scenario where time evolves forward on both copies of the field theory, introducing explicit time dependence into the system. While evolving time forward on one copy and backward on the other is a symmetry of the setup, our choice to evolve time forward on both sides breaks this symmetry, thereby generating a non-trivial time dependence in the entanglement entropy. This time dependence is reflected in the modification of the thermofield double state, where the factor $e^{-\frac{\beta\,E_n}{2}}$ is replaced by $e^{-\frac{\beta\,E_n}{2}-2iE_nt_b}$, capturing the dynamical evolution of the system \cite{Hartman:2013qma}.

The proposed holographic prescription for calculating entanglement entropy, as outlined in \cite{Santos:2024cvx,Ryu:2006bv,Hubeny:2007xt}, involves identifying an extremal codimension-two surface within the bulk spacetime. This surface must be anchored to the boundaries of the region under consideration on the conformal boundary. While this formula remains a conjecture, it has been rigorously tested in specific scenarios and has successfully passed numerous consistency checks, as discussed in \cite{Nishioka:2009un}. Despite its unproven status, the conjecture has become a cornerstone of holographic entanglement entropy studies, offering profound insights into the interplay between geometry and quantum information.

\begin{figure}[!ht]
\begin{center}
\includegraphics[width=\textwidth]{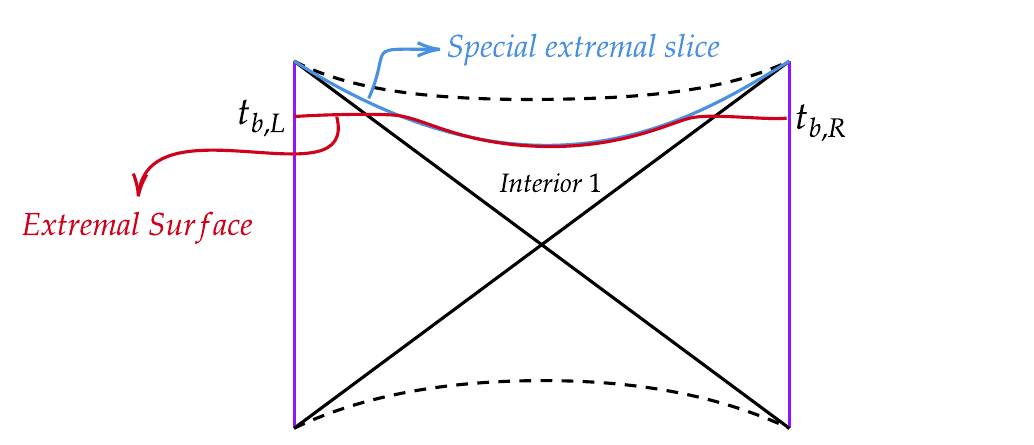}
\end{center}
\caption{The red surface responsible for computing the entanglement entropy exhibits a notable behavior for large values of $t_{b,L/R}$. In this regime, the surface closely approaches a critical spacelike surface located deep within the bulk spacetime. This critical surface plays a pivotal role in the dynamics of the system, as its proximity to the extremal surface leads to the linear dependence of the entanglement entropy on $t_b$. The linear growth reflects the accumulation of quantum correlations over time, a hallmark feature of systems with holographic duals undergoing dynamical evolution.}\label{HW1}
\label{planohwkhz}
\end{figure}
For a finite-temperature conformal field theory (CFT), we consider the region ($\mathcal{A}$) to be the half-line ( $y>0$), both in the CFT and its thermofield double. In the Euclidean signature, operator insertions in the CFT are located at ( $\text{Im}\,r=0$), while those in the thermofield double are positioned at ($\text{Im}\,r=i\beta/2$), reflecting the thermal periodicity. At ($t=0$), the ($n$)-th power of the reduced density matrix, ($\rho_{\mathcal{A}}^n$), can be represented as a Euclidean path integral on an ($n$)-sheeted cylinder with periodicity ($r\sim\,r+i\beta$). This construction involves the insertion of twist fields at the branch points, whose two-point correlation function encodes the entanglement structure. As we demonstrate, the entanglement entropy in this setup is derived from the analytic continuation of these twist field correlators in the replica limit ($n\to1$). Thus for
\begin{eqnarray}
S_{HM}=\frac{c}{6}log\,\left[\frac{2r_h\Delta_r}{r^2}\left(1+\sqrt{\Delta_L\,\Delta_R}\,\cosh\left(\frac{4\pi\,t}{\beta 
}\right)\right)\right]+\frac{c}{3}\rho_{\epsilon}.
\end{eqnarray}
we have that $S_{HM}$ for $t\gtrsim \beta$, the entanglement entropy grows linearly in time ($S_{HM}\to\frac{c}{6}t$) for $\kappa=0$. On the other hand, the region where the solution crosses the horizon and goes near the boundary has a shape that depends on the details of the geometry Fig \ref{HW1}. However, for large $t$, we have that

\begin{eqnarray}
S^{\kappa}_{HM}=\frac{c}{6}\kappa^2 \frac{Z_1}{\sqrt{Z_0^2-1}}\to\frac{c}{6}\frac{t^3}{\beta^3},\label{SHM1}
\end{eqnarray}
which increases the information area; a large fraction of CFT microstates correspond to black holes
with a smooth interior for which the construction is correct and applicable. This mapping between the black-hole interior and the boundary is state-dependent, then this neatly resolves several paradoxes about large black holes in AdS/CFT. Thus, encoded in $S^{\kappa}_{HM}$, we have a large piece lying in the interior that gives a large contribution in $t^3/\beta^3$ to the area.

\section{Conclusions and discussions}

This work demonstrates how time dependence can be incorporated to explore the black hole interior using the $\kappa$-algebra framework. In this work, we propose a specific prescription for the system's setup, which is characterized by a static geometry under time evolution. In the standard interpretation, time flows forward in one exterior region of the Penrose diagram and backward in the other. However, in our approach, we deviate from this convention by evolving time forward in both exterior regions. This modification introduces explicit time dependence into the system, allowing us to construct a simplified model for investigating thermalization in a strongly coupled conformal field theory (CFT) dual, as discussed in Refs. \cite{Hartman:2013qma, Nishioka:2009un, Hubeny:2012ry}. This approach is motivated by the observation that, while the two-sided configuration may initially seem artificial, it provides valuable insights into realistic thermalization processes. Specifically, there exists a class of black hole microstates that mimic the eternal black hole geometry outside the event horizon but do not include the second asymptotic region. These microstates correspond to time-dependent pure states in the dual CFT, which evolve and thermalize over time. By adopting this setup, we aim to bridge the gap between the idealized eternal black hole model and the more realistic dynamics of thermalization in strongly coupled systems, thereby offering a clearer understanding of the interplay between black hole physics and dual CFT dynamics.

When the entanglement entropy deviates from linear time dependence and includes third-order corrections in time ($t^3/\beta^3$), it reveals deeper insights into the dynamics of black hole thermalization and quantum information flow. In the context of holography, such corrections suggest that the thermalization process is not instantaneous but involves higher-order interactions between subsystems. These corrections could indicate the presence of subleading effects in the dual CFT, such as finite coupling corrections or deviations from perfect thermal equilibrium \cite{Faulkner:2013ica,Maldacena:2015waa}. Physically, third-order corrections to entanglement entropy imply that the growth of quantum correlations between subsystems is more intricate than a simple linear increase. This could reflect the influence of complex interactions in the black hole interior, such as the scrambling of information or the interplay between different energy scales. Additionally, these corrections may provide a window into understanding the microscopic structure of black hole horizons and the role of quantum chaos in thermalization processes.


\section*{Authors' Contributions}
All authors contributed to the study conception and design. Material preparation, data collection and analysis were performed by Fabiano F. Santos and Ankit Anand. The first draft of the manuscript was written by  Fabiano F. Santos and Ankit Anand commented on previous versions of the manuscript. All authors read and approved the final manuscript.

\section*{Data Availability}
\begin{enumerate}
\item[$\bullet$] The data supporting this work are presented in the appendix of this paper.
\item[$\bullet$] The data are in he repository https://arxiv.org/abs/2503.18919.
\item[$\bullet$] All original data for this work can be found at https://arxiv.org/abs/2503.18919.
\end{enumerate}

\section*{Conflicts of Interest}
The authors declare that there is no conflict of interest.

\section*{Ethical Considerations}
The authors have diligently addressed ethical concerns, such as informed consent, plagiarism, data fabrication, misconduct, falsification, double publication, redundancy, submission, and other related matters.

\section*{Funding}
This research did not receive any grant from funding agencies in the public, commercial, or non-profit sectors.


\section*{Acknowledgment}
A.A. is financially supported by the IIT Kanpur Institute's postdoctoral fellowship.


%
%
%
%

\end{document}